\documentclass[a4paper]{jpconf}
\usepackage{graphicx}
\begin{document}
\title{Monitoring pulsating giant stars in M33: star formation history and
chemical enrichment}

\author{Atefeh Javadi$^{1}$ and Jacco Th. van Loon$^{2}$}

\address{$^{1}$School of Astronomy, Institute for Research in Fundamental Sciences (IPM), Tehran, 19395-5531,
Iran}
\address{$^{2}$Lennard-Jones Laboratories, Keele University, ST5 5BG, UK}

\ead{atefeh@ipm.ir}

\begin{abstract}
We have conducted a near-infrared monitoring campaign at the UK InfraRed 
Telescope (UKIRT), of the Local Group spiral galaxy M 33 (Triangulum).
A new method has been developed by us to use pulsating giant stars
to reconstruct the star formation history of galaxies over cosmological
time as well as using them to map the dust production across their
host galaxies. In first Instance the central square kiloparsec of M33 was monitored 
and long period variable stars (LPVs) were identified. We give evidence of two
epochs of a star formation rate enhanced by a factor of a few. These stars 
are also important dust factories, we measure their dust production rates from a
combination of our data with {\it Spitzer} Space Telescope mid-IR photometry.
Then the monitoring survey was expanded to cover a much larger part
of M33 including spiral arms. Here we present our methodology and
describe results for the central square kiloparsec of M33
(Javadi et al.\ 2011 a,b,c, 2013) and disc
of M33 (Javadi et al.\ 2015, 2016, 2017, and in preparation).
\end{abstract}

\section{Introduction}
Luminous, cool evolved stars are powerful tracers of the underlying stellar
populations, as they stand out above all other stars especially at infrared
wavelengths and are thus the first stars that can be resolved in increasingly
distant galaxies. Asymptotic Giant Branch (AGB) stars in particular represent
stellar populations ranging in age from tens of millions of years (i.e.\ ``the
present'') to more than ten billion years (formed at redshift $\sim2$). The
cool molecular atmospheres of AGB stars lead to broad absorption troughs in
their optical spectra; while normally these are oxygen-bearing molecules such
as TiO and VO, stars in certain mass ranges dredge up carbon -- that had been
synthesized in their interiors -- and replace the absorption bands by entirely
different ones due to C$_2$ and CN. This changes their colours, generally
rendering carbon stars redder than their oxygen (M-type) counterparts. In
principle, this makes them easy to distinguish photometrically.

Triangulum offers a superb opportunity to study the structure and formation
history of a spiral galaxy, and to study various aspects of stellar evolution.
The galaxy is sufficiently massive to contain large populations of stars to be
used as tracers of star formation, and to sample brief phases of stellar
evolution such as those associated with strong mass loss on the AGB. Its disc
is seen under a more favourable viewing angle than the larger Andromeda galaxy
(M\,31), and also its more modest size ($\sim1^\circ$) makes it easier to study
it in its entirety. 
The methodology comprises three stages: [1] find stars that vary in brightness
with large amplitude (typically a magnitude) and long period (months to years),
and identify them on the basis of their colours and luminosity as cool giant stars at
the endpoints of their evolution; [2] use the fact that these stars no longer evolve
in brightness, to uniquely relate their brightness to their birth mass, and use the
birth mass distribution to reconstruct the star formation history (SFH); [3] measure
the excess infrared emission from dust produced by these stars, to quantify the
amount of matter they return to the interstellar medium in M33.

\section{Observations}
The project exploits a large observational campaign between 2003-2007, an
investment of over 100 hr on the UK InfraRed Telescope (UKIRT). In first
instance, only the central square kiloparsec ("bulge") were monitored and
analysed, with the UIST camera. The variable star catalogue and SFH were
presented in Javadi et al. (2011a,b); we found a main formation epoch of redshift
1 but with fluctuations up to the present day. As the new wide-field camera (WFCAM)
became available the campaign was expanded to cover two orders of magnitude larger
area, comprising the disc of M33 and its spiral arms.  The photometric catalogue of 
WFCAM data was presented in Javadi et al. (2015a,b). The catalogue  
comprises of 403\,734 stars, among which 4643 stars were identified as LPVs-- AGB 
stars, super--AGB stars and RSGs (Saberi et al. 2015).

\section{Star formation history}

\begin{figure}
\begin{center}
\includegraphics[width=29pc]{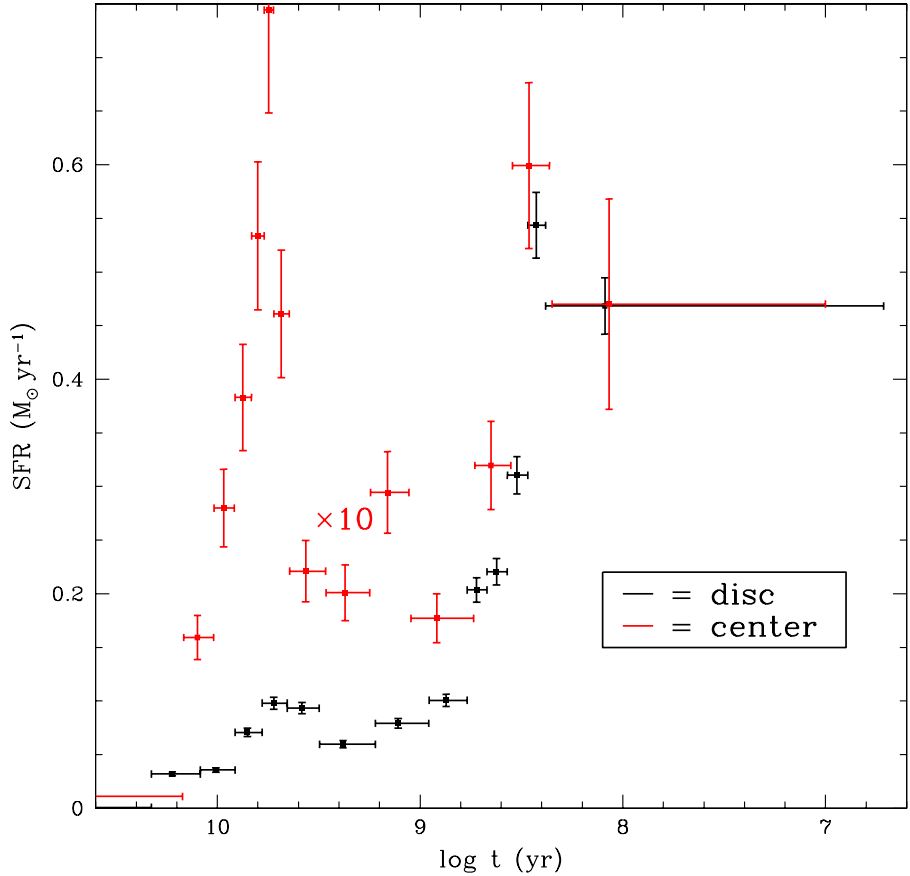}
\caption{The SFH of M\,33 in the disc (black) and in the center (red).}
\end{center}
\end{figure}
Resolved stellar populations within galaxies allow us to derive star formation histories on the
basis of colour magnitude diagram modelling, rather than from integrated light.
The LPVs are at the end-points of their evolution, and their luminosities directly reflect their
birth mass (via the core mass). Stellar evolution models provide this relation. The distribution of
LPVs over luminosity can thus be translated into the star formation history, assuming a standard
initial mass function. LPVs were formed as recently as $<$ 10 Myr ago and as long ago as $>$ 10 Gyr, so
they probe almost all of cosmic star formation. We have successfully used this new technique in M\,33
(Javadi et al.\ 2011b, 2017), the Magellanic Clouds (Rezaeikh et al.\ 2014) and NGC 147
and NGC 185 (Golshan et al., 2015, 2017), using Padova models which also provide the lifetimes of
the LPV phase. 

Fig.\ 1  presents the SFH for the center and disc of M\,33.  
In the central regions two main epochs of star formation are seen; one 
that occurred $\sim$ 4--5 Gyr ago, 
peaking around 4 Gyr ago  at a level$\sim$ 2.5 times as high as
during the subsequent couple of Gyr and 
the one that occurred $\sim$ 300 Myr--20 Myr ago with a rate $\sim$ 1.5 times higher 
than the aforementioned peak. Two main epochs of star formation are also seen in the 
disc of M\,33; one that occurred $\sim$ 6 Gyr ago and lasted $\sim$ 3 Gyr and the one
that occurred $\sim$ 250 Myr ago and lasted $\sim$ 200 Myr. More that 71 $\%$ of
stars mass in M\,33  were created during the first epoch of star formation 
and less than 13 $\%$ were created during the recent epoch of star formation.

\section{Dust production rate of evolved stars}

\begin{figure}
\begin{center}
\includegraphics[width=29pc]{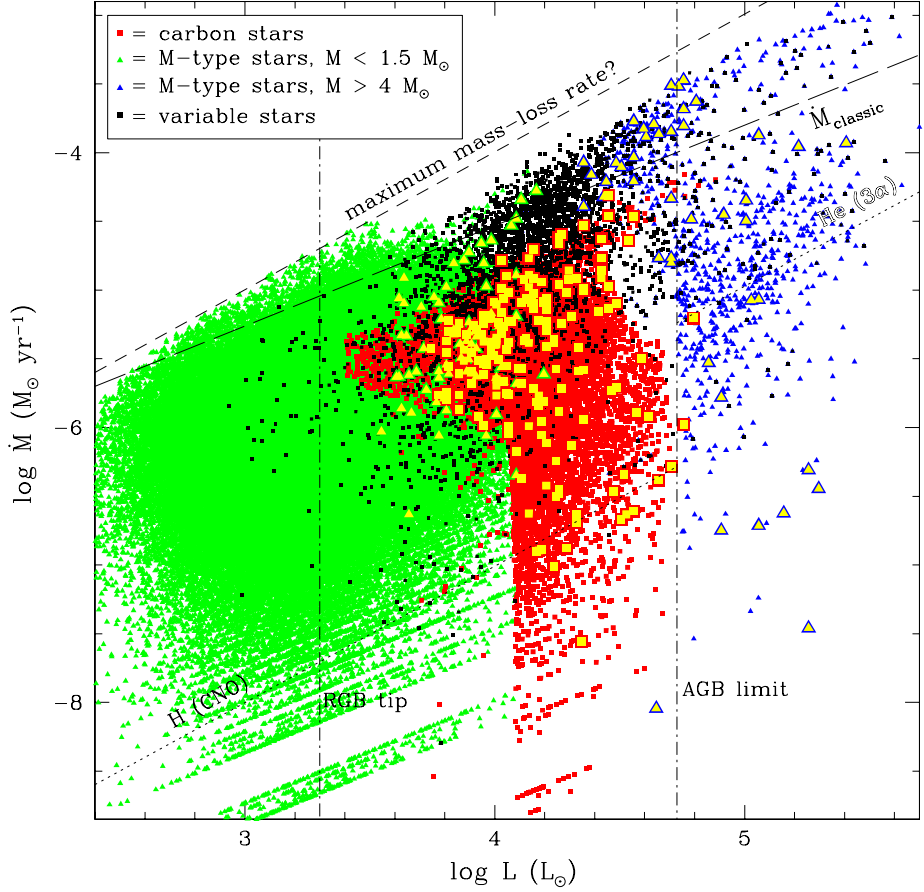}
\caption{Mass-loss vs.\ luminosity for all stars including non-variable stars. 
The blue and green triangles show the massive luminous M-type stars and low-mass 
stars (at lower luminosities), respectively. The red squares show the AGB carbon 
stars. Large yellow symbols identify the stars modelled with {\sc dusty}; other 
UKIRT variables are identified by black squares.
The vertical dash--dotted lines show the 
tip luminosity of the RGB and classical limit of the most massive AGB 
stars (Hot Bottom Burning effects are excluded). The dotted lines 
show the mass-consumption rates by shell hydrogen burning (CNO cycle) on 
the AGB and core helium burning (triple-$\alpha$ reaction) in red 
supergiants. The dashed lines show the limits to the mass-loss rate 
in dust-driven winds due to single scattering (classic) and multiple 
scattering (maximum?).}
\end{center}
\end{figure}

A majority of LPVs have been detected at mid-IR wavelengths with 
{\it spitzer} Space Telescope  (McQuinn et al.\ 2007).
Therefore, we can estimate the dust production rates across 
a galactic disc of M\,33.
We derive mass-loss rates and luminosities in two steps 
(Fig.\ 2, Paper VI in preparation). First, we use {\sc dusty} 
code (Ivezi\'c \v{Z}  \& Elitzur 1997) to model the spectral
energy distributions (SEDs) of LPVs for which
mid-IR counterparts have been identified to construct relations 
between the dust optical depth and bolometric corrections on the 
one hand, and near-IR colours on the other.
Then, by using these relations, we convert the near-IR colours 
of all red giant variables 
to luminosities and mass-loss rates.

\section{Ongoing-work and conclusions}
Currently, we are using the mass-loss rates 
to investigate the correlations between the dust production rate,
luminosity, and amplitude, and to establish a link between the 
dust return and the formation of stars within the prominent spiral arm
pattern. We will show where mass is returned, how this compares
to the gas in spiral arms and inter-arm regions, and from
this estimate gas recycle times and gas depletion in
star formation.

In conclusion, using a new technique developed by us, we showed 
how the SFH varies across M33, e.g. whether star formation
has propagated inwards or outwards through the disc and we measured the lag
between stars of different ages and the spiral arms in which they formed.

\ack{}
We are grateful for financial support by The Leverhulme Trust
under grant No. RF/4/RFG/2007/0297, by the Royal Astronomical Society,
and by the Royal Society under grant
No. IE130487.

\section*{References}

\begin{thereferences}{}
\item
     Hamedani Golshan R, Javadi A, van Loon J Th, Khosroshahi H, Saremi E
     2017 {\it MNRAS}  {\bf 466} 1764
\item
Hamedani Golshan R, Javadi A, van Loon J Th, Khosroshahi H 
2015 {\it PKAS}  {\bf 30} 169
\item
Ivezi\'c \v{Z}, Elitzur M 1997 {\it MNRAS} {\bf 287} 799
\item
Javadi A, van Loon J Th, Mirtorabi M T 2011a {\it MNRAS} {\bf 411} 263 
\item
Javadi A, van Loon J  Th, Mirtorabi M T 2011b {\it MNRAS} {\bf 414} 3394 
\item
Javadi A, van Loon J  Th, Mirtorabi M T 2011c {\it ASPC}
{\bf 445} 497
\item
Javadi A, van Loon J Th, Khosroshahi H, Mirtorabi M T 2013 {\it MNRAS} {\bf 432} 2824 
\item
Javadi A, Saberi M,  van Loon J.\ Th, Khosroshahi H, Golabatooni N,  Mirtorabi M.\ T 2015a {\it MNRAS} {\bf 447} 3973 
\item
Javadi A, van Loon J Th, Khosroshahi H 2015b {\it PKAS} {\bf 30} 355
\item
Javadi A,   van Loon J Th, Khosroshahi H 2016 {\it MmSAI} {\bf 87} 278
\item
Javadi A, van Loon J Th, Khosroshahi H, Tabatabaei F,  Hamedani Golshan R 2017 {\it MNRAS} {\bf 464} 2103 
\item
Marigo P, Girardi L, Bressan A, Groenewegen M A T, Silva L, Granato
G  L 2008 {\it A\&A} {\bf482} 883
\item
McQuinn K B W, et al 2007 {\it ApJ} {\bf 664} 850
\item
Rezaeikh S, Javadi A, Khosroshahi H, van Loon J Th 2014 {\it MNRAS} {\bf 445} 2214
\item
Saberi M, Javadi A, van Loon J Th, Khosroshahi H 2015 {\it ASPC} {\bf 497} 497
\end{thereferences}
\end{document}